\documentclass[11pt]{article}
\usepackage{eurosym}
\usepackage{amsmath}
\usepackage{amsfonts}
\usepackage{amssymb}
\usepackage{graphicx}
\usepackage{epstopdf}
\usepackage{cite}
\usepackage[usenames]{color}
\usepackage{multirow}

\providecommand{\U}[1]{\protect\rule{.1in}{.1in}}
\newcommand{\ba}{\begin{array}}
\newcommand{\ea}{\end{array}}

\newcommand{\Dsl}[1] { \setbox0=\hbox{$#1$}     
\dimen0=\wd0   \setbox1=\hbox{/} \dimen1=\wd1  \ifdim\dimen0>\dimen1        
 \rlap{\hbox to \dimen0{\hfil/\hfil}}  #1 \else \rlap{\hbox to \dimen1{\hfil$#1$\hfil}}  /  \fi  }
\newcommand{\bea}{\begin{eqnarray}}
\newcommand{\eea}{\end{eqnarray}}
\newcommand{\bs}{\boldsymbol}
\newcommand{\ns}{\Dsl{n}}

\newcommand {\nbs}{\Dsl{\bar n}}
\newcommand{\nbn}{\frac{\nbs\ns}{4}}
\newcommand{\nnb}{\frac{\ns\nbs}{4}}
\newcommand {\mb}{\bar m}

\begin{document}

\title{ {\Large  A study of relativistic corrections to $ J/\psi\rightarrow p\bar{p}$ decay} }
\author{ Nikolay Kivel \\
\textit{\ Physik-Department, Technische Universit\"at M\"unchen,}\\
\textit{James-Franck-Str. 1, 85748 Garching, Germany } }
\maketitle

\begin{abstract}
We  study  relativistic corrections  in exclusive  $S$-wave charmonium decays into proton-antiproton final state.  We calculate the NRQCD corrections to  the  dominant decay amplitude, which depend on the nucleon twist-3 light-cone distribution amplitudes only.  It is shown that in this case  the collinear factorisation is also valid beyond the leading-order approximation.  
Our numerical estimates show that  relativistic correction of relative order $v^2$ provides  large numerical impact. 

 \end{abstract}

\noindent

\vspace*{1cm}

\newpage

\section{Introduction}

In recent years, the BESII and BESIII collaborations have obtained many
accurate data on baryon-antibaryon decays of $S$-wave charmonia, see e.g.
Refs.\cite{BES:2006pax, BES:2008hwe, BESIII:2012ion,BESIII:2016ssr, BESIII:2016nix,BESIII:2017kqw, BESIII:2018flj, BESIII:2020fqg}.   
These data allows one to get information about various important hadronic 
parameters and provide an interesting possibility to study the QCD dynamics. 

There are different approaches for a description of the baryon-antybaryon
decays.  In Refs.\cite{Alekseev:2018qjg, BaldiniFerroli:2019abd} it was proposed to use a phenomenological model based on an empirical effective Lagrangian with unknown parameters, which are subject to restrictions from $SU(3)$ flavour symmetry and experimental data.

Another  way is to use effective field theory framework,  combining
nonrelativistic expansion and collinear factorisation \cite{Brodsky:1981kj, Chernyak:1983ej, Brambilla:2004wf, Chernyak:1987nv}. Appropriate effective Lagrangian is described  by  NRQCD \cite{Bodwin:1994jh, Brambilla:2004jw}  and
collinear copies of QCD \cite{Brodsky:1981kj, Chernyak:1983ej, Chernyak:1987nv}. 
 
Such approach allows one to built a systematic expansion in powers of $1/m_{c}$   describing  decay
amplitudes as a superposition of  perturbative  and universal
nonperturbative functions. The perturbative part can be computed in pQCD
 at the scale of order $m_{c}$. The nonperturbative matrix
elements are defined at soft hadronic scale $\Lambda\ll m_{c}$ . They can be limited to other processes or evaluated using non-perturbative methods such as QCD sum rules or lattice QCD.

Since the real mass of the charm quark is not large enough, a systematic description requires careful analysis of the various corrections. Existing calculations of exclusive and inclusive reactions show that in many cases radiative corrections in $\alpha_{s}(m_{c})$ and relativistic corrections can have a significant numerical effect, see for instance, reviews 
\cite{QuarkoniumWorkingGroup:2004kpm, Voloshin:2007dx}.

The first estimates for exclusive proton-antiproton decay based on the
leading-order (LO) approximation  were obtained  already long time ago in
Refs.\cite{Brodsky:1981kj,  Chernyak:1987nv}. Recently, this analysis has been extended in  Refs.\cite{Kivel:2019wjh,Kivel:2021uzl,Kivel:2022fzk}, where the power-law corrections $\sim\Lambda/m_{c}$ associated with the collinear expansion were taken into account for different amplitudes.  The collinear power corrections are associated with
the higher twist baryon light-cone distribution amplitudes (LCDAs).
It has been established that the effect of such corrections is not very strong and one can describe various channels of baryon-antibaryon decay with good accuracy. However,  relativistic and QCD radiative corrections for baryon–antibaryon  decays  have not yet been studied. 

The available data for various baryons indicate the possible presence of effects from relativistic corrections.
For example,  the so-called ``12\% rule''
\begin{equation}
Q_B=\frac{Br[\psi(2S)\rightarrow B\bar{B}]}{Br[J/\psi\rightarrow B\bar{B}]}%
\simeq\frac{Br[\psi(2S)\rightarrow e^{+}e^{-}]}{Br[J/\psi\rightarrow
e^{+}e^{-}]}\simeq0.13,
\label{12pr}
\end{equation}
must hold if the width is dominated by the LO NRQCD approximation. 
The available experimental data \cite{Zyla:2020zbs} show the following   results:   $Q_p=0.1386(3)$,  $Q_n=0.146(1)$, $Q_\Lambda=0.204(1)$, $Q_{\Sigma^0}=0.21(3)$, $Q_{\Sigma^+}=0.072(5)$ and $Q_\Xi^{+}=0.276(5)$.
Consequently, the data for baryons other than nucleon indicate a violation of this expectation.

 For  nucleon-antinucleon decays  the rule (\ref{12pr}) works  quite well, but on the other hand, the values of the polar
angular distribution coefficient $\alpha_{p}$  for ground and excited $S$-wave charmonium ($\psi(2S)\equiv \psi'$)  are very different \cite{BESIII:2012ion, BESIII:2018flj} 
\begin{equation}
\left.\alpha_{p}\right|_{J/\psi}=0.595\pm0.012,\ \ \left. \alpha_{p}\right|_{\psi^{\prime}}=1.03\pm0.06.
\end{equation}
This observation  may  also indicate  a significant  contribution of relativistic corrections.  

The purpose of this work is to study the effects of relativistic corrections in proton-antiproton decays. The technique for calculating relativistic corrections is well known. Such corrections were already
studied in  for different  exclusive decays  as $J/\psi\rightarrow e^{+}e^{-}$
 \cite{Bodwin:2002cfe, Bodwin:2007fz}, $H\rightarrow\gamma J/\psi$  \cite{Bodwin:2014bpa,Brambilla:2019fmu}, $\chi_{cJ}\rightarrow\gamma\gamma$ \cite{ Ma:2002eva,Brambilla:2006ph}  and for  exclusive production $e^{+}e^{-}\rightarrow J/\psi\eta_{c}$ \cite{Braaten:2002fi,Bodwin:2007ga}.  However, relativistic corrections to exclusive hadronic decays have not yet been considered.

The feature of hadronic  decays  is that  the hard kernels are not simple numbers but depend on light quark
momentum fractions. Corresponding decay amplitude is described  by the convolution integral over
the momentum fractions  of the hard kernel with  LCDAs. In  this case it is
more convenient to perform  the hard matching at the amplitude level.  In some
cases   the convolution  integrals have infrared divergencies  that indicates
about the violation of the collinear factorisation. This is often the case for
next-to-leading  power contributions in exclusive processes. However, for the dominant amplitude
describing $n^{3}S_{1}\rightarrow B\bar{B}$ decay, the collinear
factorisation is still applicable and this makes it possible to systematically  study  the relativistic effects. 

Our paper is organised as follows.  In Sec.~\ref{kin} we give important
definitions and kinematical notation. In  Sec.~\ref{match}  we calculate the
hard kernels  and perform the resummation of some class of relativistic
effects. After that we provide a qualitative numerical analysis for different
choices of  the nucleon LCDAs. In Sec.~\ref{disc} we discuss the obtained
results and make conclusions. In Appendix  we give important details about
the nucleon  twist-3 LCDAs and provide analytical expressions  for the hard kernels.    

\section{Kinematics and decay amplitudes }
\label{kin}

In this work we use notation from Ref.\cite{Kivel:2022fzk}. We define the kinematics of
$J/\psi(P)\rightarrow p(k)\bar{p}(k^{\prime})$ decay in the charmonium rest frame
\begin{equation}
P=M_{\psi}\omega,~\omega=(1,\vec{0}).\
\end{equation}
The outgoing momenta $k$ and $k^{\prime}$ are directed along the $z$-axis and
read%
\begin{equation}
k=(M_{\psi}/2,0,0,M_{\psi}\beta/2),~k^{\prime}=(M_{\psi}/2,0,0,-M_{\psi}%
\beta/2),\label{beta}%
\end{equation}
where $m_{N}$ is the nucleon mass and $~\beta=\sqrt{1-4m_{N}^{2}/M_{\psi}^{2}%
}$. We also define auxiliary light-cone vectors
\begin{equation}
n=(1,0,0,-1),~\ \bar{n}=(1,0,0,1),
\end{equation}
so that any four-vector $V$ can be represented as
\begin{equation}
V=V_{+}\frac{\bar{n}}{2}+V_{-}\frac{n}{2}+V_{\bot},~
\end{equation}
where $\ V_{+}\equiv(V\cdot n)=V_{0}+V_{3}$, $V_{-}\equiv(V\cdot\bar{n})=V_{0}-V_{3}$. 

The decay amplitude $J/\psi\rightarrow p\bar{p}$ is defined as
\begin{equation}
\left\langle k,k^{\prime}\right\vert ~i\hat{T}~\left\vert P\right\rangle
=(2\pi)^{4}\delta(P-k-k^{\prime})~iM,
\end{equation}
with
\begin{equation}
M=\bar{N}(k)\left\{  A_{1}%
\Dsl \epsilon_{\psi}+A_{2}\left(
\epsilon_{\psi}\right)  _{\mu}(k^{\prime}+k)_{\nu}\frac{i\sigma^{\mu\nu}%
}{2m_{N}}\right\}  V(k^{\prime})~,\label{def:M}%
\end{equation}
where
$\Dsl p=p_{\mu }\gamma^{\mu}$. The nucleon $\bar{N}(k)$ and antinucleon $V(k^{\prime})$
spinors have standard normalisation $\bar{N}N=2m_{N}$ and $\bar{V}V=-2m_{N}$.
The charmonium polarisation vector $\epsilon_{\psi}^{\mu}\equiv\epsilon_{\psi
}^{\mu}(P,\lambda)$ satisfies%
\begin{equation}
\sum_{\lambda}\epsilon_{\psi}^{\mu}(P,\lambda)\epsilon_{\psi}^{\nu}%
(P,\lambda)=-g^{\mu\nu}+\frac{P^{\mu}P^{\nu}}{M_{\psi}^{2}}.
\end{equation}

The scalar amplitudes $A_{1}$ and $A_{2}$ describe the decay process. These
amplitudes are computed in the effective field framework by expansion with
respect to small relative heavy quark velocity $v$ and   small ratio
$\lambda^{2}\sim\Lambda/m_{Q}$. The amplitude $A_{2}$ is suppressed as
\begin{equation}
A_{2}/A_{1}\sim\lambda^{2}.
\end{equation}
The decay width is conveniently described in terms of  two combinations
\begin{equation}
\mathcal{G}_{M}=A_{1}+A_{2},~\ \mathcal{G}_{E}=A_{1}+\frac{M_{\psi}^{2}%
}{4m_{N}^{2}}A_{2},
\end{equation}
and reads%
\begin{equation}
\Gamma\lbrack J/\psi\rightarrow p\bar{p}]=\frac{M_{\psi}\beta}{12\pi}\left(
\left\vert \mathcal{G}_{M}\right\vert ^{2}+\frac{2m_{N}^{2}}{M_{\psi}^{2}%
}\left\vert \mathcal{G}_{E}\right\vert ^{2}\right)  .\label{width}%
\end{equation}
The ratio $\left\vert \mathcal{G}_{E}\right\vert /\left\vert \mathcal{G}%
_{M}\right\vert $ can be measured through the angular behaviour of the cross
section $e^{+}e^{-}\rightarrow J/\psi\rightarrow p\bar{p}$, see {\it e.g.}
Ref.\cite{Barnes:2007ub}.  Available data indicate that the contribution with the
amplitude $\mathcal{G}_{E}$ is  quite small (about $10\%$) and the value of
width is dominated by the amplitude $\left\vert \mathcal{G}_{M}\right\vert^{2}$ .    

To the leading-order approximation
\begin{equation}
\mathcal{G}_{M}\simeq A_{1}^{\text{lo}}+\mathcal{O}(\alpha_{s})+\mathcal{O}%
(v^{2})+\mathcal{O}(\lambda^{2}),
\end{equation}
where in the \textit{rhs} we indicate various possible corrections. These are 
next-to-leading (NLO) order contribution with respect to $\alpha_{s}$, relativistic
corrections of relative order $v^{2}$ and power corrections associated with
the higher twist light-cone distribution amplitudes, respectively.  The NLO
radiative QCD corrections and relativistic corrections  are not known yet.
The power suppressed contribution  have been estimated in Ref.\cite{Kivel:2022fzk} and
corresponding  numerical effect is moderate,  about $25\%$.  
In this paper we want  to study the relativistic corrections to the amplitude  $A_{1}^{\text{lo}}$.  

\begin{figure}[ptb]%
\centering
\includegraphics[width=3.50in]{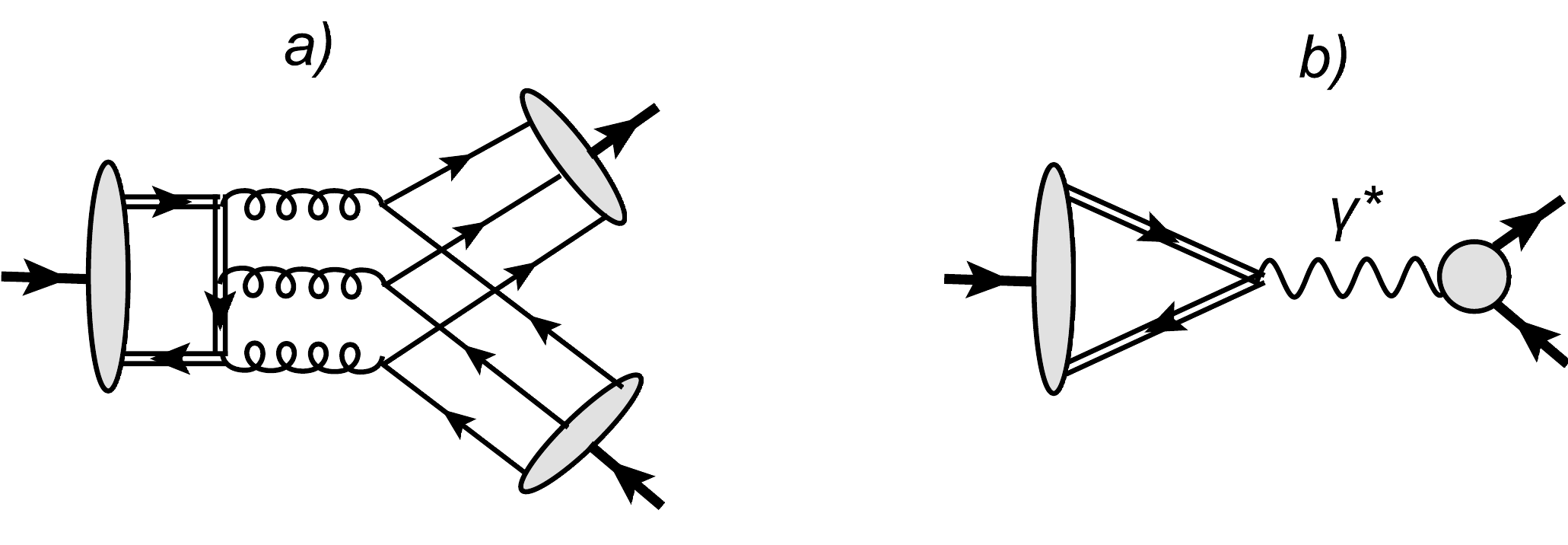}
\caption{ Hadronic $a)$ and electromagnetic $b)$ contributions to the decay amplitudes. }
\label{figure}
\end{figure}

The amplitude  $A_{1}^{\text{lo}}$ is described by the sum of two different  contributions: the hard and electromagnetic one,
which correspond to the annihilation into three gluons or  one hard  photon as shown in Fig.\ref{figure}. The electromagnetic amplitude is relatively  small, however  it can provide numerical impact through the interference with
QCD amplitude Ref.\cite{Kivel:2022fzk}.  In this paper, for simplicity, 
the electromagnetic contribution is not considered. 

The leading-order factorisation formula for the amplitude $A_{1}^{\text{lo}}$
reads \cite{Brodsky:1981kj,Chernyak:1983ej}%
\begin{equation}
A_{1}^{(0)}=\frac{\sqrt{2M_{\psi}}\left\langle 0\right\vert O\left\vert
J/\psi\right\rangle }{~m_{c}^{2}}\frac{~f_{N}^{~2}}{m_{c}^{4}}~(\pi\alpha
_{s})^{3}~\frac{10}{81}~J_{0},\label{A1:LO}%
\end{equation}
where  the NRQCD matrix element is defined in the standard way%
\begin{equation}
\left\langle 0\right\vert O\left\vert J/\psi\right\rangle =\left\langle
0\right\vert \chi^{\dag}\ \boldsymbol{\sigma}\cdot\boldsymbol{\epsilon
}(\lambda)\psi\left\vert J/\psi(\lambda)\right\rangle =\sqrt{\frac{N_{c}}%
{2\pi}}R_{10}(0),
\end{equation}
where $R_{10}(0)$ is the radial charmonium wave function at the origin.
The constant $f_{N}$ has the dimension of the square of the mass and describes 
the normalization of the matrix element of the nucleon.
The decay amplitude is sensitive to nucleon structure that is described by the dimensionless collinear convolution integral $J_{0}$,
which reads
\begin{align}
J_{0}  & =\frac{1}{4}\int Dx_{i}\int Dy_{i}\ \frac{1}{x_{1}x_{2}x_{3}}\frac
{1}{y_{1}y_{2}y_{3}}\frac{y_{1}x_{3}}{D_{1}D_{3}}\label{def:J0}\\
& \ \ \ \ \ \ \ \ \ \ \ \ \times\left\{  \varphi_{3}(y_{123})~\varphi
_{3}(x_{123})+\frac{1}{2}\left(  \varphi_{3}(y_{123})+\varphi_{3}%
(y_{321})\right)  \left(  \varphi_{3}(x_{123})+\varphi_{3}(x_{321})\right)
\right\}  ,\nonumber
\end{align}
where $D_{i}=x_{i}+y_{i}-2x_{i}y_{i}$  and we used short notation for the
twist-3 nucleon LCDA $\varphi_{3}(x_{123})\equiv\varphi_{3}(x_{1},x_{2}%
,x_{3})$. This nonperturbative function depends on the quark momentum
fractions $0<x_{i}<1$ and the factorisation scale, which is not shown explicitly.
The  light-cone fractions $x_{i}$ satisfy the momentum conservation
condition $x_{1}+x_{2}+x_{3}=1$. Therefore the measure of the convolution
integrals in Eq.(\ref{def:J0}) includes the $\delta$-function
\begin{equation}
Dx_{i}=dx_{1}dx_{2}dx_{3}\delta(1-x_{1}-x_{2}-x_{3}).
\end{equation}

The nucleon LCDA $\varphi_{3}$ is defined as the matrix element of a 3-quark
operator and is well known in the literature, see {\it e.g.} Refs.\cite{Chernyak:1983ej,Braun:2000kw}.
 For a convenience, the definition and some important details are given in Appendix. 

Our task is to calculate the relativistic corrections associated
with the higher order matrix elements in NRQCD%
\begin{equation}
\left\langle 0\right\vert O_{n}\left\vert J/\psi\right\rangle =\left\langle
0\right\vert \chi^{\dag}\boldsymbol{\sigma}\cdot\boldsymbol{\epsilon}%
(\lambda)\left(  -\frac{i}{2}\overleftrightarrow{\boldsymbol{D}}\right)
^{2n}\psi\left\vert J/\psi(\lambda)\right\rangle ,\ \ \label{def:On}%
\end{equation}
where $\left\langle 0\right\vert O_{0}\left\vert J/\psi\right\rangle
\equiv\ \left\langle 0\right\vert O\left\vert J/\psi\right\rangle $.
 Usually, it is convenient to describe the contribution of these matrix elements  introducing  the
ratio%
\begin{equation}
\left\langle \boldsymbol{q}^{2n}\right\rangle =\frac{\left\langle 0\right\vert
O_{n}\left\vert J/\psi\right\rangle }{\left\langle 0\right\vert O_{0}%
\left\vert J/\psi\right\rangle }.
\label{def:q2}
\end{equation}
The set of the operators in Eq.(\ref{def:On}) does not represent  the complete set of the all
possible NRQCD operators.  The operator $O_2$ provides the correct description of the relativistic
correction of relative order $v^2$. But to describe relativistic corrections of a higher order, many more different operators are used, for example, colour-octet ones. However, resumming the contributions of all orders associated with the subset (\ref{def:q2}) is useful because it allows one to study the convergence of the nonrelativistic expansion.

The value of the NLO  matrix element (\ref{def:q2})  can be
exactly  found   with the help of Kapustin-Gremm relation Ref.\cite{Gremm:1997dq}
\begin{equation}
\left\langle 0\right\vert O_{1}\left\vert J/\psi\right\rangle =m_{c}%
E_{J/\psi}\ \left\langle 0\right\vert O_{0}\left\vert J/\psi\right\rangle ,
\end{equation}
where $E_{J/\psi}$ denotes the binding energy.  This gives
\begin{equation}
\left\langle \boldsymbol{q}^{2}\right\rangle =m_c E_{J/\psi}.
\end{equation}
In Ref.\cite{Bodwin:2006dn} it was found that the  higher order operators satisfy approximate relation
\begin{equation}
\left\langle \boldsymbol{q}^{2n}\right\rangle =\left\langle \boldsymbol{q}%
^{2}\right\rangle ^{n}+\mathcal{O}(v^{2}).\label{q2n}%
\end{equation}
Neglecting the higher order corrections in the \textit{rhs} of Eq.(\ref{q2n}) allows one to
resumm the power series of the  higher order terms with the matrix elements
(\ref{def:On}) to all orders. Such resummation includes those relativistic
corrections that are contained in the quark-antiquark quarkonium wave function
in the leading potential model for the wave function \cite{Bodwin:2007ga}. 

The  resummation  of the relativistic corrections associated with the matrix elements (\ref{def:q2})
 is already considered for various  processes in Refs.\cite{Bodwin:2002cfe,Bodwin:2007fz,Bodwin:2007ga} and this technique can be easily adapted for calculations of exclusive hadronic decays. 

\section{Matching and  all order resummation }
\label{match}

In order to calculate  the decay amplitude we use the covariant spin-projector
technique \cite{Kuhn:1979bb}.   The generalisation of this  technique to  higher orders in
$v^{2}$ is considered in Refs.\cite{Bodwin:2002cfe,Bodwin:2007fz} .   Here we briefly  describe the NRQCD
matching and clarify  some  features of the present calculation.

Our task is to compute the hard coefficient functions in front of higher order
operators (\ref{def:On}). The matching can be conveniently done using
$Q\bar{Q}$-state as initial state instead of charmonium state. The
corresponding  heavy quark and antiquark have the  momenta $p$ and $p^{\prime
}$, respectively
\begin{equation}
p=(E,\boldsymbol{q}),\ p^{\prime}=(E,-\boldsymbol{q}),\ \ \ E=\sqrt{m_{Q}%
^{2}+\boldsymbol{q}^{2}}.
\end{equation}
Then total and relative momenta reads%
\begin{equation}
P=p+p^{\prime},\ \ q=\frac{1}{2}(p-p^{\prime}).
\end{equation}
In the rest frame
\begin{equation}
P=(2E,\boldsymbol{0}),\ \ q=(0,\boldsymbol{q}).
\end{equation}

Let the perturbative amplitude is described as \
\begin{equation}
A_{QQ}=\bar{v}_{Q}(p^{\prime},s^{\prime})\ \hat{A}_{QQ}\ u_{Q}(p,s),
\end{equation}
where $\bar{v}_{Q}$ and $u_{Q}$ denote heavy quark spinors.  The  amplitude
projected onto $S$-wave state of heavy quark-antiquark $Q\bar{Q}(^{3}S_{1})$
reads
\begin{equation}
A_{QQ}(^{3}S_{1})=\ \text{Tr}\left[  \Pi_{1}\hat{A}_{QQ}(x_{i},y_{i})\right]
,\label{A3S1}%
\end{equation}
 where the spin-triplet  projector $\Pi_{1}$ is given by \cite{Bodwin:2002cfe}:%
\begin{equation}
\Pi_{1}=\frac{-1}{2\sqrt{2}E(E+m_Q)}\left(  \frac{1}{2}\Dsl{P}+\Dsl q+m_Q\right)
\frac{\Dsl P+2E}{4E}\Dsl \epsilon\left(  \frac{1}{2}\Dsl P-\Dsl q-m_Q\right)
\otimes\frac{\mathbf{1}}{\sqrt{N_{c}}}.
\end{equation}
This progector is relativistically normalised%
\begin{equation}
\text{Tr[}\Pi_{1}\Pi_{1}^{\dag}\text{]}=4E^{2}.
\end{equation}
In order to project the amplitude onto state with  $L=0$  the expression in
Eq.(\ref{A3S1})  must be averaged  over the angles of the momentum $\boldsymbol{q}$
\begin{equation}
\left.  A_{QQ}(\boldsymbol{q}^{2})\right\vert _{^{3}S_{1},\ L=0}=\bar{A}%
_{QQ}(\boldsymbol{q}^{2})=\frac{1}{4\pi}\int d\Omega\ \text{Tr}\left[  \Pi
_{1}\hat{A}_{QQ}(x_{i},y_{i})\right]  .\label{A3S1L0}%
\end{equation}

In NRQCD the amplitude reads%
\begin{equation}
A_{\text{NRQCD}}=\sqrt{2M_{\psi}}\sum_{n\geq0}c_{n}\ \left\langle 0\right\vert
O_{n}\left\vert J/\psi\right\rangle ,\label{Anrqcd}%
\end{equation}
where \ $\sqrt{2M_{\psi}}$ arises from relativistic normalisation. The
coefficients $c_{n}$ can be obtained from the amplitude  (\ref{A3S1L0}) using
that
\begin{equation}
\left\langle 0\right\vert O_{n}\left\vert Q\bar{Q}\right\rangle =\sqrt{2N_{c}%
}2E\ \boldsymbol{q}^{2n},
\end{equation}
and
\begin{equation}
\bar{A}_{QQ}(\boldsymbol{q}^{2})=\sum_{n}c_{n}\left\langle 0\right\vert
O_{n}\left\vert Q\bar{Q}\right\rangle =\sum_{n}c_{n}\sqrt{2N_{c}%
}2E\ \boldsymbol{q}^{2n}.
\end{equation}
Using this one finds%
\begin{equation}
c_{n}=\frac{1}{\sqrt{2N_{c}}}\frac{1}{n!}\frac{\partial}{\partial
\boldsymbol{q}^{2n}}\frac{\bar{A}_{QQ}(\boldsymbol{q}^{2})}{2E}.
\end{equation}
Substituting this into  Eq.(\ref{Anrqcd}) and using Eq.(\ref{q2n}) \ one finds%
\bea
A_{\text{NRQCD}}\simeq\sqrt{2M_{\psi}}\left\langle 0\right\vert O\left\vert
J/\psi\right\rangle \sum_{n}\frac{1}{n!}\left\langle \boldsymbol{q}%
^{2}\right\rangle ^{n}\frac{\partial^n}{\partial\boldsymbol{q}^{2n}}\frac
{\bar{A}_{QQ}(^{3}S_{1})}{\sqrt{2N_{c}}2E}
 \\
 =\sqrt{2M_{\psi}}\left\langle
0\right\vert O\left\vert J/\psi\right\rangle \frac{\bar{A}_{QQ}(\left\langle
\boldsymbol{q}^{2}\right\rangle )}{\sqrt{2N_{c}}2E}.
\eea

\begin{figure}[h]%
\centering
\includegraphics[width=3.50in]{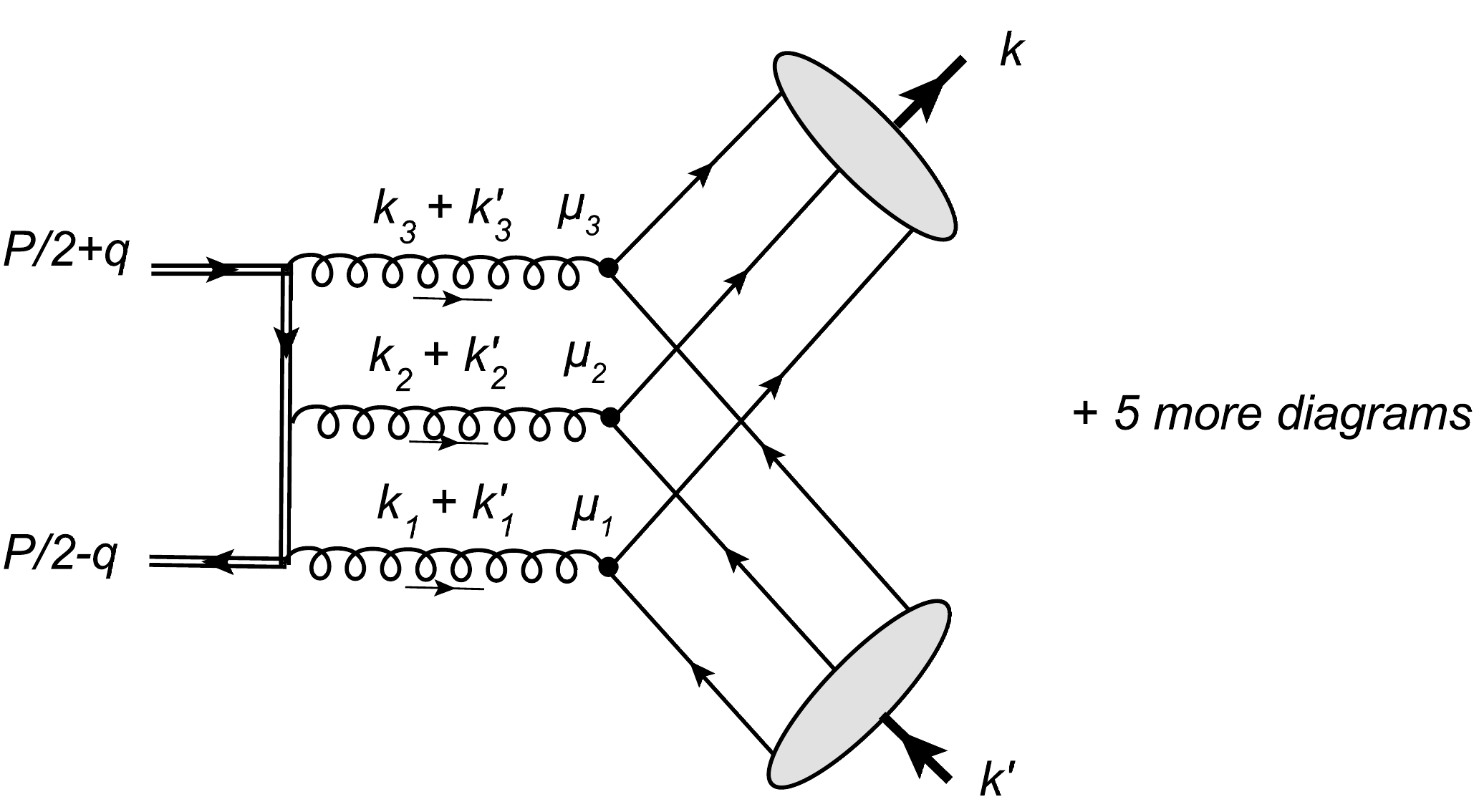}
\caption{ The diagrams, which describe $Q\bar{Q}\rightarrow p\bar{p}$  subprocess.  }
\label{hard_diagms}
\end{figure}

The specific of our calculation is that the amplitude $\bar{A}_{QQ}(^{3}%
S_{1})$ also includes the nonperturbative matrix elements describing the
couplings with the nucleons. The amplitude $Q\bar{Q}\rightarrow p\bar{p}$ \ is desribed by the diagrams in
Fig. \ref{hard_diagms}. 
Each diagram can be described as consisting of a heavy quark subdiagram $ \Gamma_{Q}$ describing the $Q\bar{Q}$ annihilation into three gluons, and a residual part describing the transition of three virtual gluons to the $p\bar{p}$ state. The expression for each diagram  can be written in the following form
\bea
A_{QQ}^{icj}=\int Dx_{i}\int Dy_{i}\ \ \Delta_{1}^{\mu_{1}\mu_{1}^{\prime}%
}\Delta_{2}^{\mu_{2}\mu_{2}^{\prime}}\Delta_{3}^{\mu_{3}\mu_{3}^{\prime}%
}\ B_{\mu_{1}\mu_{2}\mu_{3}}(x_{i},y_{i})\ 
  \bar{v}_{Q}(p^{\prime},s^{\prime
}) \Gamma_{Q}^{\mu_{i}^{\prime}\mu_{c}^{\prime}\mu_{j}^{\prime}%
}  u_{Q}(p,s),
\label{def:AQQ}
\eea
where $\bar{v}_{Q}$ and $u_{Q}$ denote heavy quark spinors, $\Delta_{i}%
^{\mu_{i}\mu_{i}^{\prime}}$ denotes the gluon propagator (we use Feynman gauge
and do not write the colour indices for simplicity)%
\begin{equation}
\Delta_{i}^{\mu_{i}\mu_{i}^{\prime}}=\frac{(-i)g^{\mu_{i}\mu_{i}^{\prime}}%
}{(k_{i}+k_{i}^{\prime})^{2}},\ k_{i}=x_{i}k,\ \ k_{i}^{\prime}=y_{i}%
k^{\prime},
\end{equation}
where the proton and antiproton momenta reads
\begin{equation}
k\simeq E\ n,\ \ k^{\prime}\simeq E\ \bar{n},\ \ k^{2}=k^{\prime2}=0.
\end{equation}

The heavy quark subdiagram reads
\begin{equation}
\Gamma_{Q}^{\mu_{i}^{\prime}\mu_{c}^{\prime}\mu_{j}^{\prime}}(x_{i}%
,y_{i})= i^2(ig)^3\frac{ \gamma^{\mu_{i}} (-\Dsl P/2+\Dsl q+\Dsl k_{i}%
+\Dsl k_{i}^{\prime}+m_{Q})\gamma^{\mu_{c}}(\Dsl P/2+\Dsl q-\Dsl k_{j}-\Dsl k_{j}^{\prime}+m_{Q})\gamma^{\mu_{j}}}
{\left[(k_{i}+k_{i}^{\prime}-P/2+q)^{2}-m_{Q}^{2}\right]  \left[  (P/2+q-k_{j}%
-k_{j}^{\prime})^{2}-m_{Q}^{2}\right]  }.\label{Ajci}%
\end{equation}
The set of indices $\{\mu_{i},\mu_{c},\mu_{j}\}$ describes six possible
transpositions of the indices $\{\mu_{1},\mu_{2},\mu_{3}\}$, which correspond
to the different diagrams: \ $\{\mu_{1},\mu_{2},\mu_{3}\}$, $\{\mu_{3},\mu
_{2},\mu_{1}\}$, $\{\mu_{2},\mu_{1},\mu_{3}\}$, $\{\mu_{3},\mu_{1},\mu_{2}\}$,
\ $\{\mu_{1},\mu_{3},\mu_{2}\}$, $\{\mu_{2},\mu_{3},\mu_{1}\}$. Therefore each
diagram can be coded by the set $icj$ and the sum of these diagrams gives the
total result. 

The term $B_{\mu_{1}\mu_{2}\mu_{3}}(x_{i},y_{i})$ in Eq.(\ref{def:AQQ}) describes the trace, which
occur after contractions of the Dirac indices $\alpha_{i}$ and $\alpha_{i}^{\prime}$ 
of the proton and antiproton light-cone  matrix elements with  the $\gamma$-matrices of the  quark-gluon vertices
\begin{align}
& \left\langle 0\right\vert \left[  O_{\text{tw3}}\right]  _{\alpha
_{1}^{\prime}\alpha_{2}^{\prime}\alpha_{3}^{\prime}}~\left\vert \bar
{p}(k^{\prime})\right\rangle ~\left[  \left(  ig\right)  \gamma_{\mu_{1}%
}\right]  _{\alpha_{1}^{\prime}\alpha_{1}}\left[  \left(  ig\right)
\gamma_{\mu_{2}}\right]  _{\alpha_{2}^{\prime}\alpha_{2}}\left[  \left(
ig\right)  \gamma_{\mu_{3}}\right]  _{\alpha_{3}^{\prime}\alpha_{3}%
}~\left\langle 0\right\vert \left[  O_{\text{tw3}}\right]  _{\alpha_{1}%
\alpha_{2}\alpha_{3}}~\left\vert p(k)\right\rangle \\
& =\ \text{FT}\left[  B_{\mu_{1}\mu_{2}\mu_{3}\rho}(x_{i},y_{i})\right]  .
\label{def:B}
\end{align}
The symbol \textquotedblleft FT\textquotedblright\ denotes Fourier
transformations, which occur from the light-cone matrix elements (the 
definition of the light-cone operator $O_{\text{tw3}}$ is given in
Appendix \ref{LCDA}).   The nucleon matrix elements are  described by the convenient
combinations of  LCDAs  $V_{1}$,$\ A_{1}$ and $T_{1}$, which \ are defined as
\begin{align}
V_{1}(x_{123})  & =\frac{1}{2}\left(  \varphi_{3}(x_{123})+\varphi_{3}%
(x_{213})\right)  ,\\
A_{1}(x_{123})  & =\frac{1}{2}\left(  \varphi_{3}(x_{123})-\varphi_{3}%
(x_{213})\right)  ,
\end{align}%
\begin{equation}
T_{1}(x_{123})=\frac{1}{2}\left(  \varphi_{3}(x_{132})+\varphi_{3}%
(x_{231})\right)  .
\end{equation}

 The $B$-term (\ref{def:B}) is universal for all diagrams.  The  calculation of the Dirac
contractions yields%
\begin{equation}
B_{\mu_{1}\mu_{2}\mu_{3}}(x_{i},y_{i})=\bar{N}_{\bar{n}} \gamma_{\bot}^{\rho}{V}_{n}
 (ig)^{3}\frac{1}{4}(k\cdot k^{\prime})\ ~T_{\mu_{1}\mu_{2}\mu
_{3}\rho}(x_{i},y_{i}),
\end{equation}
with
\begin{align}
T_{\mu_{1}\mu_{2}\mu_{3}\rho}(x_{i},y_{i})  & =\ g_{\mu_{1}\mu_{2}}^{\bot
}g_{\rho\mu_{3}}^{\bot}\left\{  V_{1}(x_{123})V_{1}(y_{123})+A_{1}%
(x_{123})A_{1}(y_{123})\right\}  \label{T123rho}\\
& -i\varepsilon_{\mu_{1}\mu_{2}}^{\bot}i\varepsilon_{\mu_{3}\rho}^{\bot
}\left\{  V_{1}(y_{123})A_{1}(x_{123})+A_{1}(y_{123})V_{1}(x_{123})\right\}
\nonumber\\
& +G_{\mu_{1}\mu_{2}\alpha\sigma}^{\bot}G_{\rho\mu_{3}}^{\bot\alpha\sigma
}~T_{1}(y_{123})T_{1}(x_{123}).\nonumber
\end{align}
where we used the following short notations%
\begin{equation}
g_{\alpha\beta}^{\bot}\ =g_{\alpha\beta}-\frac{1}{2}\left(  n_{\alpha}\bar
{n}_{\beta}+n_{\beta}\bar{n}_{\alpha}\right)  ,\ \ i\varepsilon_{\alpha\beta
}^{\bot}=\frac{1}{2}i\varepsilon_{\alpha\beta\sigma\lambda}n^{\sigma}\bar
{n}^{\lambda},
\end{equation}%
\begin{equation}
G_{\alpha\beta\sigma\rho}^{\bot}=g_{\alpha\sigma}^{\bot}g_{\beta\rho}^{\bot
}+g_{\alpha\rho}^{\bot}g_{\beta\sigma}^{\bot}-g_{\sigma\rho}^{\bot}%
g_{\alpha\beta}^{\bot}.\
\end{equation}
For the projected nucleon spinors $\bar{N}(k)$ and are $V(k^{\prime})$  we used%
\begin{equation}
\bar{N}(k)\nnb=\bar{N}_{ \bar{n}},\ \ 
\nnb V(k^{\prime})=V_{n}.
\end{equation}

Therefore  the expression for the amplitude (\ref{A3S1L0})  reads\
\begin{equation}
i\bar{A}_{QQ}(\boldsymbol{q}^{2})=\frac{1}{2}\frac{1}{4\pi}\int d\Omega
\sum_{\{icj\}}\int Dx_{i}\int Dy_{i} \ T_{\mu_{1}\mu_{2}\mu_{3}} 
 \Delta_{1}^{\mu_{1}\mu_{1}^{\prime}}\Delta_{2}^{\mu_{2}\mu_{2}^{\prime}}\Delta_{3}^{\mu_{3}\mu_{3}^{\prime}}
 \text{Tr}\left[  \Pi_{1}\hat
{A}_{QQ}^{\mu_{i}^{\prime}\mu_{c}^{\prime}\mu_{j}^{\prime}}\right]  ,
\label{AbarQQ}
\end{equation}
\ where we also show explicitly the symmetry factor $1/2$.  Calculating the
trace and contracting the indices we obtain the following result\footnote{We used  the package Feyncalc \cite{Kublbeck:1992mt}.}
\begin{equation}
\bar{A}_{QQ}(\boldsymbol{q}^{2})=\sqrt{2N_{c}}\frac{2}{m_{Q}}\frac{f_{N}^{2}%
}{(m_{Q}^{2}+\boldsymbol{q}^{2})^{2}}\ (\pi\alpha_{s})^{3}\ \frac{10}{81}%
\frac{m_{Q}+E}{2m_{Q}}\ J(\boldsymbol{q}^{2}/m_{Q}^{2}),
\label{barAQQ}
\end{equation}
where the dimensionless collinear integral reads%
\begin{equation}
J(\boldsymbol{v}^{2})=\frac{1}{32}\frac{1}{f_{N}^{2}}\int\ \frac{Dx_{i}}%
{x_{1}x_{2}x_{3}}\int Dy_{i}\frac{1}{y_{1}y_{2}y_{3}}\left\{  \frac
{A(x_{i},y_{i})}{D_{1}D_{3}}+\frac{B(x_{i},y_{i})}{D_{1}D_{2}}+\frac
{C(x_{i},y_{i})}{D_{2}D_{3}}\right\}  .\ \label{def:J}%
\end{equation}
The integrand in Eq.(\ref{def:J}) \ has the following structure%
\begin{equation}
D_{i}=x_{i}+y_{i}-2x_{i}y_{i}=x_{i}(1-y_{i})+y_{i}(1-x_{i})\geq0.
\end{equation}
The functions $A,B$ and $C$ also depend on $\boldsymbol{v}^{2}=\boldsymbol{q}%
^{2}/m_{Q}^{2}$, which is not  explicitly shown for simplicity. They can be
presented in the following way%
\begin{align}
A(x_{123},y_{123})  & =\sum_{i=0}^{4}A_{k}(x_{123},y_{123})\ I_{k}%
(1,3),\ \label{def:Ak}\\
B(x_{123},y_{123})  & =\sum_{i=0}^{4}B_{k}(x_{123},y_{123})\ I_{k}%
(1,2),\label{def:Bk}\\
\ C(x_{123},y_{123})  & =\sum_{i=0}^{4}C_{k}(x_{123},y_{123})\ I_{k}%
(2,3).\label{def:Ck}%
\end{align}

These expressions include the angular integrals $I_{k}$ which appears due to
the integration over $d\Omega=d\varphi\ d\cos\theta$. The polar integration
can be easily computed and the remaining  azimuth integrals read
\begin{equation}
I_{k}[ij]=\frac{1}{2}\int_{-1}^{1}d\eta\ \frac{|\boldsymbol{v}|^{k}\eta^{k}%
}{\left[  1+|\boldsymbol{v}|a_{i}\eta\right]  \left[  1-|\boldsymbol{v}%
|a_{j}\eta\right]  },\ \label{def:Ik}%
\end{equation}
where we used $\ \eta\equiv\cos\theta,\ \ |\boldsymbol{v}|=|\boldsymbol{q}%
|/m_{Q}$ and convenient short notation%
\begin{equation}
a_{i}=\frac{m_Q}{E}\frac{x_{i}-y_{i}}{D_{i}}<1.\label{def:ai}%
\end{equation}
The numerator in (\ref{def:Ik}) \ arises from the trace\ and contractions in
Eq.(\ref{AbarQQ}). The denominators $\left[  1+|\boldsymbol{v}|a_{i}%
\eta\right]  \left[  1-|\boldsymbol{v}|a_{j}\eta\right]  $ \ occur from the
heavy quark \ propagators. The angular dependence occurs from the scalar
products%
\begin{equation}
(qk)=-Eq_{z}=-E|\boldsymbol{q}|\cos\theta,\ \ (qk^{\prime})=Eq_{z}%
=E|\boldsymbol{q}|\cos\theta.
\end{equation}

From the definition (\ref{def:Ik}) it follows that%
\begin{equation}
I_{0}\sim\mathcal{O}(1),\ \ I_{1,2}\sim\mathcal{O}(\boldsymbol{v}%
^{2}),\ \ \ I_{3,4}\sim\mathcal{O}(\boldsymbol{v}^{4}).
\end{equation}
If one neglects by the contribution form the denominators setting
\ $a_{i}=a_{j}=0$ then
\begin{equation}
I_{0}[ij]=1,\ I_{1}[ij]=0,\ I_{2}[ij]=\frac{\boldsymbol{v}^{2}}{3}%
,\ I_{3}[ij]=0,\ I_{4}=\frac{\boldsymbol{v}^{2}}{5}.
\end{equation}
One can also easily find that%
\begin{equation}
I_{k}[ji]=(-1)^{k}I_{k}[ij].
\end{equation}
These integrals can be computed analytically \ that gives%
\begin{align}
I_{0}[ij]  & =\frac{1}{2}\frac{1}{a_{i}+a_{j}}\left(  \ln\frac
{1+|\boldsymbol{v}|a_{i}}{1-|\boldsymbol{v}|a_{i}}+\ln\frac{1+|\boldsymbol{v}%
|a_{j}}{1-|\boldsymbol{v}|a_{j}}\right)  \nonumber\\
& =\frac{1}{a_{i}+a_{j}}\sum_{n=0}^{\infty}\frac{|\boldsymbol{v}|^{n}}%
{n+1}\left(  a_{j}^{n+1}+(-1)^{n}a_{i}^{n+1}\right)  \frac{1}{2}\left[
1+(-1)^{n}\right]  ,\label{I0:res}%
\end{align}
and for $k>0$
\begin{equation}
I_{k}[ij]=\frac{1}{2}\frac{1}{a_{i}+a_{j}}\frac{(-1)^{k}}{a_{i}^{k-1}}\left(
\frac{1}{|\boldsymbol{v}|a_{i}}\ln\frac{1+|\boldsymbol{v}|a_{i}}%
{1-|\boldsymbol{v}|a_{i}}-\sum_{l=0}^{k-1}|\boldsymbol{v}|^{l}a_{i}%
^{l}(-1)^{l}\frac{1+(-1)^{l}}{l+1}\right)
\end{equation}%
\begin{equation}
+\frac{1}{2}\frac{1}{a_{i}+a_{j}}\frac{1}{a_{j}^{k-1}}\left(  \frac
{1}{|\boldsymbol{v}|a_{i}}\ln\frac{1+|\boldsymbol{v}|a_{i}}{1-|\boldsymbol{v}%
|a_{i}}-\sum_{l=0}^{k-1}|\boldsymbol{v}|^{l}a_{j}^{l}\frac{1+(-1)^{l}}%
{l+1}\right)
\end{equation}%
\begin{equation}
=\frac{|\boldsymbol{v}|^{k}}{a_{i}+a_{j}}\sum_{n=0}^{\infty}\frac
{|\boldsymbol{v}|^{n}}{n+1+k}\left(  a_{j}^{n+1}+(-1)^{n}a_{i}^{n+1}\right)
\frac{1}{2}\left[  1+(-1)^{n+k}\right]  .\label{Ik:res}%
\end{equation}

The coefficients $A_{k},B_{k}$ and $C_{k}$ in Eqs.(\ref{def:Ak})-(\ref{def:Ck}%
) \ depend on the nucleon LCDAs and their general structure reads
($X=\{A,B,C\}$)%
\begin{align}
X_{k}(x_{123},y_{123})  & =\left[  X_{k}\right]  _{VV}\left\{  V_{1}%
(x_{123})V_{1}(y_{123})+A_{1}(x_{123})A_{1}(y_{123})\right\}  
\nonumber \\
& +\left[  X_{k}\right]  _{AV}\left\{  A_{1}(x_{123})V_{1}(y_{123}%
)+V_{1}(x_{123})A_{1}(y_{123})\right\} 
\nonumber  \\
& +\left[  X_{k}\right]  _{TT}\ T_{1}(x_{123})T_{1}(y_{123}),
\label{defXk}
\end{align}
where coefficients $\left[  X_{k}\right]  _{VV,AV,TT}$ are polynomial
functions of the momentum fractions $x_{i}$ and $y_{i}$.  Their analytical expressions
are presented in Appendix~\ref{appXk}. 

The convolution integral $J$ in Eq.(\ref{def:J}) is well defined that can
be easily understood  taking into account the general behaviour of the LCDAs
\begin{equation}
Z(x_{123})=x_{1}x_{2}x_{3}\times (\text{polynomial function in }%
x_{i}),\ Z=\left\{  V_1,A_1,T_1\right\}  .
\end{equation}

Using the expression in Eq.(\ref{barAQQ}) one finds  the  NRQCD amplitude $A_{1}$ defined in
Eq.(\ref{def:M}) %
\begin{equation}
\left(  A_{1}\right)  _{\text{NRQCD}}\ =\frac{\sqrt{2M_{\psi}}\left\langle
0\right\vert O\left\vert J/\psi\right\rangle }{m_{c}^{2}\sqrt{1+\left\langle
\boldsymbol{v}^{2}\right\rangle }}\frac{f_{N}^{2}}{m_{c}^{4}(1+\left\langle
\boldsymbol{v}^{2}\right\rangle )^{2}}\ (\pi\alpha_{s})^{3}\ \frac{10}%
{81}\left(  1+\sqrt{1+\left\langle \boldsymbol{v}^{2}\right\rangle
}\right) \frac{1}{2} \ J(\left\langle \boldsymbol{v}^{2}\right\rangle ).\label{A1NRQCD}%
\end{equation}
In the limit $\left\langle \boldsymbol{v}%
^{2}\right\rangle \rightarrow0$ this result reproduces the well known
leading-order approximation in Eq.(\ref{A1:LO}). The obtained expression
depends on the power of  heavy quark mass $m_{c}^{-6}$ as required by the
scale properties of the amplitude. The charm mass in the  factor
$\sqrt{2M_{\psi}}$ is usually calculated  as $M_{\psi}\simeq2m_{c}%
\sqrt{1+\left\langle \boldsymbol{v}^{2}\right\rangle }$.  The factor
$(1+\left\langle \boldsymbol{v}^{2}\right\rangle )^{-2}$ is naturally provided by
the hard propagators in the diagrams.  The essential part of the relativistic
corrections is included in the convolution integral, which depends on
the twist-3 LCDAs, which describe the nonperturbtive overlaps with outgoing
nucleons. Therefore the  total effect of the relativistic corrections
also depends on the nonpertubative structure of nucleon.  

To see  numerical impact of the obtained relativistic corrections we
need to compute the convolution integral $J$. This integral is not simple and, in the general case, can only be calculated numerically. In order to understand better the dependence of the
integral on the model of the LCDA we consider two different models. As an example of
 relatively simple case,  consider the  asymptotic LCDA,  which is defined as 
\begin{equation}
V_{1}^{as}=T_{1}^{as}=120\ f_{N}\ x_{1}x_{2}x_{3},\ \ \ A_{1}^{as}%
=0.\ \ \label{VATas}%
\end{equation}
Appropriate analysis allows better understanding  numerical values of various convolution integrals. 
For a more realistic description  we use ABO-model \cite{Anikin:2013aka}, which provides a reliable description of various data. 

For our consideration, we use $m_{c}=1.4$ GeV for the pole heavy quark mass and use the estimate for the binding energy $J/\psi$ from the Ref.\cite{Bodwin:2007fz}%
\begin{equation}
E_{J/\psi}=0.306\text{ GeV},
\end{equation}
that gives 
\begin{equation}
\left\langle \boldsymbol{v}^{2}\right\rangle _{J/\psi}\simeq
0.225\ .\label{v2Jpsi}%
\end{equation}
The value $\left\langle \boldsymbol{v}^{2}\right\rangle _{\psi(2S)}%
\equiv\left\langle \boldsymbol{v}^{2}\right\rangle _{\psi^{\prime}}$ can be
estimated \ as%
\begin{equation}
\left\langle \boldsymbol{v}^{2}\right\rangle _{\psi^{\prime}}=\frac{E_{\psi'}%
}{m_{c}}\simeq\frac{M_{\psi^{\prime}}-M_{J/\psi}+E_{J/\psi}}{(M_{J/\psi
}-E_{J/\psi})/2}=0.64,\label{v2psip}%
\end{equation}
which is qute large due to the large difference $M_{\psi^{\prime}}-M_{J/\psi
}\simeq589$ MeV$.$

\subsection{Relativistic corrections with asymptotic nucleon LCDA}

The asymptotic DA (\ref{VATas}) is not a realistic model, but the use of this approximation makes it possible to simplify analytical expressions and study the properties of convolution integrals.  At first step we also  calculate the correction  of relative order  $v^{2}$ only. 

The results in Eqs. (\ref{I0:res}) and (\ref{Ik:res})  yield
\begin{equation}
I_{0}[ij]=1+\left\langle \boldsymbol{v}^{2}\right\rangle \frac{1}{3}\left(
a_{i}^{2}-a_{i}a_{j}+a_{j}^{2}\right)  +\mathcal{O}(v^{4})\simeq1+\left\langle
\boldsymbol{v}^{2}\right\rangle I_{0}^{(1)}[ij],
\end{equation}%
\begin{align}
I_{1}[ij]  & =\frac{\left\langle \boldsymbol{v}^{2}\right\rangle }{3}%
(a_{j}-a_{i})+\mathcal{O}(v^{4})\simeq I_{1}^{(1)}[ij],\ \ \\
\ \ I_{2}[ij]  & =\frac{\left\langle \boldsymbol{v}^{2}\right\rangle }%
{3}+\mathcal{O}(v^{4}),\ \ I_{3,4}\sim\mathcal{O}(v^{4}),
\end{align}
where  $a_{i}$ is defined in Eq.(\ref{def:ai}). Therefore  the contributions
with  $I_{3,4}$ can be neglected. \ It is  conveninet  to define
\begin{equation}
J_{as}(\left\langle \boldsymbol{v}^{2}\right\rangle )=\frac{120^{2}}{32}%
\sum_{k=0}^{4}J_{as}^{(k)}(\left\langle \boldsymbol{v}^{2}\right\rangle ),
\end{equation}
where the coefficient in front of the sum  is choosen in order to get a
convenient normalisation.  In what follow we are going to discuss the values of the 
integrals
\begin{equation}
J_{as}^{(k)}(\left\langle \boldsymbol{v}^{2}\right\rangle )=\frac{1}%
{120^{2}f_{N}^{2}}\int\ \frac{Dx_{i}}{x_{1}x_{2}x_{3}}\int\frac{Dy_{i}}%
{y_{1}y_{2}y_{3}}\left\{  \frac{A_{k}^{as}I_{k}[13]}{D_{1}D_{3}}+\frac
{B_{k}^{as}I_{k}[12]}{D_{1}D_{2}}+\frac{C_{k}^{as}I_{k}[23]}{D_{2}D_{3}%
}\right\} ,
\label{def:Jask}%
\end{equation}
and their  dependence on the parameter $\left\langle
\boldsymbol{v}^{2}\right\rangle $. 

For the coefficients $A_{k}^{as}$ we get%
\begin{align}
A_{k}^{as}  & =\left[  A_{k}\right]  _{VV}\left\{  V_{1}^{as}(x_{123}%
)V_{1}^{as}(y_{123})\right\}  +\left[  A_{k}\right]  _{TT}\ T_{1}^{as}%
(x_{123})T_{1}^{as}(y_{123})\\
& =f_{N}^{2}\ 120^{2}(x_{1}x_{2}x_{3})(y_{1}y_{2}y_{3})\ \left(  \left[
A_{k}\right]  _{VV}+\left[  A_{k}\right]  _{TT}\right)  =f_{N}^{2}%
\ 120^{2}(x_{1}x_{2}x_{3})(y_{1}y_{2}y_{3})\ \bar{A}_{k}^{as},
\end{align}
with
\begin{equation}
\bar{A}_{k}^{as}=\left[  A_{k}\right]  _{VV}+\left[  A_{k}\right]  _{TT},
\end{equation}
and similarly for $B_{k}$ and $C_{k}$. Substituting this  in
Eq.(\ref{def:Jask}) we obtain \
\begin{equation}
J_{as}^{(k)}\simeq\int Dx_{i}\int Dy_{i}\left\{  \frac{\bar{A}_{k}^{as}%
I_{k}[13]}{D_{1}D_{3}}+\frac{\bar{B}_{k}^{as}I_{k}[12]}{D_{1}D_{2}}+\frac
{\bar{C}_{k}^{as}I_{k}[23]}{D_{2}D_{3}}\right\}
\end{equation}%
\begin{equation}
=\int Dx_{i}\int Dy_{i}\frac{I_{k}[12]}{D_{1}D_{2}}\left\{  \bar{B}_{k}%
^{as}+\hat{P}_{23}\bar{A}_{k}^{as}+(-1)^{k}\hat{P}_{13}\bar{C}_{k}%
^{as}\right\}  ,
\end{equation}
where the transposition operator $\hat{P}_{ij}$ intechanges the arguments
$\{x_{i},y_{i}\}\leftrightarrow\{x_{j},y_{j}\}$, for instance
\begin{equation}
\hat{P}_{23}f(x_{123},y_{123})=f(x_{132},y_{132}).
\end{equation}
Substituting the explicit expressions for the coefficients $\bar{A}_{k}^{as}$,
$\bar{B}_{k}^{as}$ and $\bar{C}_{k}^{as}$ and neglecting the higher order
terms  we get%
\begin{align}
\sum_{k=0}^{2}J_{as}^{(k)}(\left\langle \boldsymbol{v}^{2}\right\rangle )  &
=\int Dx_{i}\int Dy_{i}\ \frac{1}{D_{1}D_{2}}\left\{  24x_{1}y_{2}%
-8\left\langle \boldsymbol{v}^{2}\right\rangle +24\left\langle \boldsymbol{v}%
^{2}\right\rangle x_{1}y_{2}(I_{0}^{(1)}[12]+I_{1}^{(1)}[12])\right. 
\nonumber \\
& \ \ \ \ \ \ \ \ \ \ \ \ \ \ \ \ \ \ \ \ \ \ \ \ \ \ \ \ \ \ \ \ \ \ \left.
+\left\langle \boldsymbol{v}^{2}\right\rangle \left(  2x_{1}y_{2}-\frac{1}%
{6}\left(  x_{1}+x_{2}+y_{1}+y_{2}\right)  \right)  \right\}  .
\end{align}
The integrals, which enter in these formulae  can be easily calculated
numerically 
\begin{equation}
\text{LO:}\int Dx_{i}\int Dy_{i}\ \frac{\ x_{1}y_{2}}{D_{1}D_{2}%
}=0.140.\label{lo}%
\end{equation}%
\begin{equation}
\text{NLO}:\ 1^{st}=\int Dx_{i}\int Dy_{i}\ \frac{1}{D_{1}D_{2}}%
=2.029,\label{nlo1st}%
\end{equation}%
\begin{align}
\ \text{NLO: }2^{nd}  & =\int Dx_{i}\int Dy_{i}\ \frac{x_{1}y_{2}}{D_{1}D_{2}%
}(I_{0}^{(1)}[12]+I_{1}^{(1)}[12])=0.012,\\
\text{NLO: }3^{d}  & =\int Dx_{i}\int Dy_{i}\ \frac{1}{D_{1}D_{2}}\left(
2x_{1}y_{2}-\frac{1}{6}\left(  x_{1}+x_{2}+y_{1}+y_{2}\right)  \right)
=-0.121.
\end{align}
This gives%
\begin{align}
\sum J_{as}^{(k)}(\left\langle \boldsymbol{v}^{2}\right\rangle )  &
=24\ast0.140\ -8\left\langle \boldsymbol{v}^{2}\right\rangle \ast\left(
2.029\right)  _{1st}+24\left\langle \boldsymbol{v}^{2}\right\rangle \left(
0.012\right)  _{2nd}+\left\langle \boldsymbol{v}^{2}\right\rangle \left(
-0.121\right)  _{3d}\label{sumJask}\\
& =3.36_{\text{lo}}-16.232\left\langle \boldsymbol{v}^{2}\right\rangle
_{1st}+0.167\left\langle \boldsymbol{v}^{2}\right\rangle _{2nd+3d}%
=3.36-16.40\left\langle \boldsymbol{v}^{2}\right\rangle .
\end{align}
\ Substituting  the numerical values for $\left\langle \boldsymbol{v}%
^{2}\right\rangle _{J/\psi}$ (\ref{v2Jpsi}) and $\left\langle \boldsymbol{v}%
^{2}\right\rangle _{\psi^{\prime}}$ (\ref{v2psip}) we obtain%
\begin{equation}
\sum J_{as}^{(k)}(\left\langle \boldsymbol{v}^{2}\right\rangle _{J/\psi
})=-0.33,
\end{equation}%
\begin{equation}
\sum J_{as}^{(k)}(\left\langle \boldsymbol{v}^{2}\right\rangle _{\psi^{\prime
}})=-7.14.
\end{equation}
These  results show that the relativistic correction is negative and
provides  about $100\%$ and $300\%$ numerical effect for $1S$ and $2S$-states,
respectively.  To better understand the origin of such a large numerical impact, consider the structure of the expression in the equation. (\ref{sumJask}).

 From this result it follows that  relativistic correction is dominated by
the terms which appear from the integral (\ref{nlo1st}). This integral is
about an order of magnitude larger than the leading-order integral (\ref{lo}).
This has the natural mathematical explanation: the both integrands in
Eqs.(\ref{lo}) and (\ref{nlo1st})  are positive, but the leading-order
integrand in (\ref{lo}) has in the numerator  small  factor $x_{1}y_{2}$
 that reduces  the value of  the integral (\ref{lo})
comparing to one in (\ref{nlo1st}).   The actual values  of the  NRQCD
 parameter $\left\langle \boldsymbol{v}^{2}\right\rangle $ for charmonium
states can not compensate the  numerical enhancement of the
subleading integral.  This  indicates about the relative  smallness of the charm quark
mass. 

The mechanism of the enhancement of the subleading integral does not depend
on the model of  LCDA of the nucleon.  The structure of  leading- and
subleading integrals  is closely related to  the structure of the
numerators of the Feynman diagrams and can be  interpreted as a specific
feature of the  hard perturbative subprocess.  In order to see this
consider the  leading-order approximation with $q=0$, then  the trace in
Eq.(\ref{AbarQQ})  reads%
\bea
\text{Tr}\left[  \Pi_{1}^{(\text{lo})}\Gamma_{Q}^{\mu_{i}^{\prime}\mu
_{c}^{\prime}\mu_{j}^{\prime}}(x_{i},y_{i})\right]  \sim\text{Tr}\left[
\left(   \Dsl \omega+1\right)  \Dsl \epsilon_{\psi} \gamma_{\bot}^{\mu_{i}%
}(- \Dsl P/2+m_{Q}+ \Dsl k_{i}+ \Dsl k_{i}^{\prime})\gamma_{\bot}^{\mu_{c}}
\right.
\nonumber \\ 
\left.
\times ( \Dsl P/2+ \Dsl k_{j}%
+ \Dsl k_{j}^{\prime}+m_{Q})\gamma_{\bot}^{\mu_{j}}\right]  \label{TrPi0}%
\eea
where we used expression from Eq.(\ref{Ajci}) and
\begin{equation}
\ \Pi_{1}^{(\text{lo})}=\frac{-1}{2\sqrt{2}}\left(  \Dsl \omega+1\right)
\Dsl \epsilon_{\psi}.
\end{equation}
Notice that \ all Dirac matrices with  open Lorentz indices  in
Eq.(\ref{TrPi0}) are transvers, because \ of \ contraction with transverse
tensor $T_{\mu_{1}\mu_{2}\mu_{3}\rho}$ in Eq.(\ref{T123rho}). Remind, that the
heavy qaurk velocity
\begin{equation}
\omega^{\mu}=\delta_{\mu0}=\frac{1}{2}(n^{\mu}+\bar{n}^{\mu}).
\end{equation}
If one neglects  collinear momenta $k_{i}$, $k_{i}^{\prime}$ or $k_{j}$,
$k_{j}^{\prime}$ in the trace Eq.(\ref{TrPi0}) then the trace vanishes, for
instance
\bea
\text{Tr}\left[  \left(  \Dsl \omega+1\right)   \Dsl  \epsilon_{\psi}%
\gamma_{\bot}^{\mu_{i}}(- \Dsl P/2+m_{Q})\gamma_{\bot}^{\mu_{c}}( \Dsl P/2+ \Dsl k_{j}%
+ \Dsl k_{j}^{\prime}+m_{Q})\gamma_{\bot}^{\mu_{j}}\right]
\\
 =m_{Q}\text{Tr}\left[  \left(  \Dsl \omega+1\right)  \Dsl  \epsilon_{\psi
}\ \gamma_{\bot}^{\mu_{i}}(- \Dsl \omega+1)\gamma_{\bot}^{\mu_{c}}(\ldots
)\gamma_{\bot}^{\mu_{j}}\right] 
 \\
 =m_{Q}\text{Tr}\left[  \underline{\left(   \Dsl  \omega+1\right)
(- \Dsl  \omega+1)} \Dsl  \epsilon_{\psi}\ \gamma_{\bot}^{\mu_{i}}\gamma_{\bot
}^{\mu_{c}}(\ldots)\gamma_{\bot}^{\mu_{j}}\right]  =0,
\eea
where we used $[ \Dsl \omega,\gamma_{\bot}]=[ \Dsl \omega, \Dsl  \epsilon_{\psi}]=0$.
Therefore the nontrivial result is obtained only from the term with collinear
momenta in both quark propagators%
\begin{equation}
\text{Tr}\left[  \Pi_{1}^{(\text{lo})}\hat{A}_{QQ}^{\mu_{i}^{\prime}\mu
_{c}^{\prime}\mu_{j}^{\prime}}(x_{i},y_{i})\right]  \sim\text{Tr}\left[
\left(  \Dsl  \omega+1\right)   \Dsl  \epsilon_{\psi}\gamma_{\bot}^{\mu_{i}%
}( \Dsl k_{i}+ \Dsl k_{i}^{\prime})\gamma_{\bot}^{\mu_{c}}( \Dsl k_{j}+ \Dsl k_{j}^{\prime}%
)\gamma_{\bot}^{\mu_{j}}\right]  \sim x_{i}y_{j}+x_{j}y_{i}.
\end{equation}
This explains  the structure of the leading-order integral in Eq.(\ref{lo}). 

The calculation of the subleading correction involves the relative momentum $q$
that introduces the traces with two insertions of  $q$  instead of collinear
momenta, for instance
\begin{equation}
\text{Tr}\left[  \Pi_{1}\Gamma_{Q}^{\mu_{i}^{\prime}\mu_{c}^{\prime}\mu
_{j}^{\prime}}(x_{i},y_{i})\right]  \sim\text{Tr}\left[  \left(
 \Dsl  \omega+1\right)   \Dsl  \epsilon_{\psi}\gamma_{\bot}^{\mu_{i}} \Dsl  q%
\gamma_{\bot}^{\mu_{c}} \Dsl  q\gamma_{\bot}^{\mu_{j}}\right]  +\ \ldots\ .
\end{equation}
Such subleading contributions give terms of relative order $v^{2}$ but the
corresponding integral  has no  momentum fractions in the numerator, see 
Eq.(\ref{nlo1st}).  As a result,  the large numerical contributions are
generated  by the integrals  $J_{as}^{(0)}$ and $J_{as}^{(2)}$
only, which can only get appropriate contributions from the trace and
contractions in Eq.(\ref{AbarQQ}).  Parametrically such integrals are
suppressed by small velocity, however, in reality such  suppression is not
sufficiently  strong  in order to provide a small numerical correction with
respect to the leading-order contribution. 

The enhanced  integral also takes place  in the relative order $v^{4}$, but  the corresponding numerical
effect is already suppressed by additional factor  $\left\langle
\boldsymbol{v}^{2}\right\rangle$ and for $J/\psi$ such a contribution can be
estimated  at about  $20-30\%$ . But for the excited state such a contribution still
remains  quite large because of large value of $\left\langle \boldsymbol{v}^{2}\right\rangle _{\psi^{\prime}}$.   

 In order to study the more general situation and the possible numerical effects
of higher order terms in $v^{2}$,  we present in Table~\ref{tabJas} the results for the
integrals $J_{as}^{(k)}$, which are calculated with the  resummed
relativistic corrections. Remind, that the leading-order approximation is
given by the integral $J_{as}^{(0)}(\left\langle \boldsymbol{v}^{2}\right\rangle =0)=3.36.$
\begin{table}[th]
\caption{
Numerical  results for  integrals $J_{as}^{(k)}$ for $J/\psi$ and $\psi'$,  upper and bottom tables, respectively. 
}
\label{tabJas}%
\centering
\begin{tabular}
[c]{|c|c|c|c|c|c|c|}\hline
 $\left\langle \boldsymbol{v}^{2}\right\rangle _{J/\psi}=0.225$  & $J_{as}^{(0)}$ & $J_{as}^{(1)}$ & $J_{as}^{(2)}$ & $J_{as}^{(3)}$ & $J_{as}^{(4)}$ &
$\sum J_{as}^{(k)}$\\\hline
$\text{NLO}$ & $-1.84$ & $-0.18$ & $1.89$ & $-$ & $-$ & $-0.13$\\\hline
$\text{ all orders }$ & $-1.00$ & $-0.19$ & $1.66$ & $-0.01$ & $0.13$ &
$0.59$\\\hline
\end{tabular}
\\[4mm]%
\begin{tabular}
[c]{|c|c|c|c|c|c|c|}\hline
 $\left\langle \boldsymbol{v}^{2}\right\rangle _{\psi^{\prime}}=0.64$ 
 & $J_{as}^{(0)}$ & $J_{as}^{(1)}$ & $J_{as}^{(2)}$ & $J_{as}^{(3)}$ &
$J_{as}^{(4)}$ & $\sum J_{as}^{(k)}$\\\hline
$\text{NLO}$ & $-11.68$ & $-0.50$ & $5.37$ & $-$ & $-$ & $-6.81$\\\hline
$\text{all orders}$ & $-6.78$ & $-0.69$ & $4.53$ & $-0.14$ & $0.98$ &
$-2.10$\\\hline
\end{tabular}
\end{table}

The largest numerical effects are provided by the NLO corrections in
integrals $J_{as}^{(0,2)}$ for  the same reason as  discussed above.
 The numerical effect from the the resummation is most visible for
$J_{as}^{(0)}$ but corresponding effects  are not sufficiently large and  do not change 
 the qualitative picture discussed for NLO approximation. Therefore we can conclude
that relativistic expansion for $J/\psi$  is well convergent. The large
numerical impact is only generated by the correction of relative order $v^{2}$
and associated with the numerical enhancement of the NLO convolution integral.

In the Table~\ref{tabJas} (bottom part) we also show the results for the $2S$-state, which has much larger
relativistic corrections. In this case the general structure remains quite similar: the largest numerical effect
is provided by the NLO contribution in   $J_{as}^{(0,2)}$ but  the higher
order power corrections also make a significant numerical  contribution,
reducing the total sum by a factor of three.  This clearly illustrates  the  expected  conclusion: the relativistic expansion for excited charmonium has very large relativistic corrections  and converges rather slowly.

\subsection{Relativistic corrections with realistic nucleon LCDA }

The value of the collinear integral  also depends on the  model of the  LCDA
 $\varphi_{3}$, which describes a distribution of the quark longitudinal
momenta at zero transverse separation. 
To illustrate the possible effect of the nucleon structure, we calculate in this section the convolution integrals with a realistic LCDA model.  For that purpose we use the
ABOI-model from Ref.\cite{Anikin:2013aka}. This model gives a reliable description of the
electromagnetic form factor data within the light-cone sum rules \cite{Balitsky:1986st} and
also provides a good description of $J/\psi\rightarrow p\bar{p}$ decay data in
the leading-order approximation \cite{Kivel:2022fzk}. The expression for the model reads
\bea
&&\varphi_{3}^{\text{ABO}}(x_{i})=120x_{1}x_{2}x_{3}\left\{  1+\varphi
_{10}\mathcal{P}_{10}(x_{i})+\varphi_{11}\mathcal{P}_{11}(x_{i}) 
\right. 
\nonumber \\
&& \ \ \ \ \ \ \ \ \ \ \ \ \ \ \ \ \ \ \ \left.
 \phantom{ 120x_{1}x_{2}x_{3} }+\varphi
_{20}\mathcal{P}_{20}(x_{i})+\varphi_{21}\mathcal{P}_{21}(x_{i})+\varphi
_{22}\mathcal{P}_{22}(x_{i})\right\}  ,
\label{phi3ABO}
\eea
where the orthogonal polynomials $\mathcal{P}_{ij}(x_i)$ read
\begin{equation}
\mathcal{P}_{10}(x_{i})=21(x_{1}-x_{3}),~~\mathcal{P}_{11}(x_{i}%
)=7(x_{1}-2x_{2}+x_{3}),\label{P1i}%
\end{equation}%
\begin{align}
&  \mathcal{P}_{20}(x_{i})=\frac{63}{10}\left[  3(x_{1}-x_{3})^{2}%
-3x_{2}(x_{1}+x_{3})+2x_{2}^{2}\right]  ,~\\
&  \mathcal{P}_{21}(x_{i})=\frac{63}{2}(x_{1}-3x_{2}+x_{3})(x_{1}-x_{3}),\\
&  \mathcal{P}_{22}(x_{i})=\frac{9}{5}\left[  x_{1}^{2}+9x_{2}(x_{1}%
+x_{3})-12x_{1}x_{3}-6x_{2}^{2}+x_{3}^{2}\right]  .
\label{P2i}%
\end{align}
The  moments $\varphi_{ij}\equiv$ $\varphi_{ij}(\mu)$ are multiplicatively
renormalisable, more details  about properties of the polynomials
$\mathcal{P}_{ij}$ and about the evolution of the moments can be found in
Ref.\cite{Braun:1999te}. In Appendix \ref{LCDA} we also provide some useful details. In our calculation we
fix the relatively low normalisation scale $\mu^{2}=1.5$ GeV following to
Ref.\cite{Kivel:2022fzk}.  Then the values of the twist-3 moments read
\begin{equation}
\varphi_{10}=0.051,\ \varphi_{11}=0.052,\ \ \varphi_{20}=0.078,\ \ \varphi
_{21}=-0.028,\ \ \varphi_{22}=0.179.
\end{equation}
 Again, we define the integral as  the sum 
\begin{equation}
J_{\text{ABO}}(\left\langle \boldsymbol{v}^{2}\right\rangle )=\frac{120^{2}%
}{32}\sum_{k=0}^{4}J_{\text{ABO}}^{(k)}(\left\langle \boldsymbol{v}%
^{2}\right\rangle ),
\end{equation}
where the integrals \ $J_{\text{ABO}}^{(k)}$ are defined as%
\begin{equation}
J_{\text{ABO}}^{(k)}(\left\langle \boldsymbol{v}^{2}\right\rangle )=\frac
{1}{120^{2}f_{N}^{2}}\int\ \frac{Dx_{i}}{x_{1}x_{2}x_{3}}\int\frac{Dy_{i}%
}{y_{1}y_{2}y_{3}}\left\{  \frac{A_{k}^{\text{ABO}}I_{k}[13]}{D_{1}D_{3}%
}+\frac{B_{k}^{\text{ABO}}I_{k}[12]}{D_{1}D_{2}}+\frac{C_{k}^{\text{ABO}}%
I_{k}[23]}{D_{2}D_{3}}\right\}  .
\end{equation}

The value of the leading order integral reads 
\begin{equation}
J_{\text{ABO}}^{(0)}(\left\langle \boldsymbol{v}^{2}\right\rangle =0)=5.08.
\end{equation}
The numerical results for the different integrals $J_{\text{ABO}}^{(k)}$ for $J/\psi$ and $\psi^{\prime}$ are
presented in Table~\ref{tabJABO}.
\begin{table}[th]
\caption{Numerical  results for the integrals $J_{\text{ABO}}^{(k)}$ for $J/\psi$ and $\psi'$, upper and bottom tables, respectively. \newline}%
\label{tabJABO}%
\centering%
\begin{tabular}
[c]{|c|c|c|c|c|c|c|}\hline
$\left\langle \boldsymbol{v}^{2}\right\rangle _{J/\psi}=0.225$ &
$J_{\text{ABO}}^{(0)}$ & $J_{\text{ABO}}^{(1)}$ & $J_{\text{ABO}}^{(2)}$ &
$J_{\text{ABO}}^{(3)}$ & $J_{\text{ABO}}^{(4)}$ & $\sum J_{\text{ABO}}^{(k)}%
$\\\hline
$\text{NLO}$ & $-5.18$ & $-0.27$ & $3.66$ & $-$ & $-$ & $-1.79$\\\hline
$\text{all orders}$ & $-3.65$ & $-0.35$ & $3.30$ & $-0.03$ & $0.22$ &
$-0.51$\\\hline
\end{tabular}
\\[4mm]%
\begin{tabular}
[c]{|c|c|c|c|c|c|c|}\hline
$\  \left\langle \boldsymbol{v}^{2}\right\rangle _{\psi^{\prime}}=0.64\  $ &
$J_{\text{ABO}}^{(0)}$ & $J_{\text{ABO}}^{(1)}$ & $J_{\text{ABO}}^{(2)}$ &
$J_{\text{ABO}}^{(3)}$ & $J_{\text{ABO}}^{(4)}$ & $\sum J_{\text{ABO}}^{(k)}%
$\\\hline
$\text{NLO}$ & $-24.09$ & $-0.75$ & $10.42$ & $-$ & $-$ & $-14.42$\\\hline
$\text{all orders}$ & $-15.69$ & $-1.39$ & $9.31$ & $-0.30$ & $1.78$ &
$-6.29$\\\hline
\end{tabular}
\end{table}

The qualitative picture remains the same as described above,  the sum of the
integrals $J_{\text{ABO}}^{(0)}$+$J_{\text{ABO}}^{(2)}$  provides the
largest numerical impact but  the values of the all integrals are  larger.
 We can conclude that  the described mechanism of the large  numerical effect 
 also works for the realistic model of LCDAs.  In case of  $J/\psi$  the total sum is negative, which indicates 
 that the negative relativistic correction is somewhat larger than the LO contribution. 

Additional numerical impact is also provided by  $\left\langle \boldsymbol{v}%
^{2}\right\rangle $-dependence of the coefficient in front of of the
convolution integral in Eq.(\ref{A1NRQCD}). The dominant numerical effect is
provided by the factor  $1/(1+\left\langle \boldsymbol{v}^{2}\right\rangle
)^{2}$,  which occurs from the hard propagators.  It seems that this factor
can be understood as an indication that  the charmonium mass $M_{\psi}%
^{2}\simeq4m_c^{2}(1+\left\langle \boldsymbol{v}^{2}\right\rangle )$ is a
more natural scale for the hard gluon propagators in the Feynman diagrams. The
corresponding numerical effect is smaller in comparison with one in the
convolution integral discussed above.  For $J/\psi$ the modification of
the coefficient in Eq.(\ref{A1NRQCD}) gives a reduction about  $37\%$. For
$\psi^{\prime}$, a similar effect is already  $67\%$.

\section{Discussion}
\label{disc}

We presented  the  first study of  relativistic corrections  in exclusive
$S$-wave charmonium decays into proton-antiproton final state.  The
relativistic corrections are calculated within the NRQCD  and collinear
factorisation frameworks.  We only consider  the  helicity conserving
amplitude $A_{1}$, which provides the dominant numerical contribution to the
decay width. In this case  the baryon non-perturbative light-cone matrix
elements depend on the twist-3 LCDAs only and the  formula for the amplitude with
relativistic corrections includes the collinear convolution integral,  which
is free from any infrared singularities.   

The result obtained provides a full correction of the relative order $v^{2}$, and also includes the summation of all orders of relativistic corrections associated with the quark-antiquark-quarkonia wave function in the potential model  \cite{Bodwin:2007ga}.   

 The largest numerical impact  is provided by the NLO relativistic correction with the matrix element, which can be computed using  equation of motion  and is proportional to the binding energy.  Due to a specific structure  of the LO and NLO hard scattering kernels the value of  the NLO collinear integral is about an order of magnitude larger  than the
value of  LO one.  The charmonium parameter $\left\langle \boldsymbol{v}^{2}\right\rangle $ is not small enough to compensate for this effect, and the magnitude of the resulting relativistic correction is numerically of the same order as the LO contribution.
For the case of  $J/\psi$ the  resulting NLO correction is negative and  therefore,  the two
contributions almost cancel out. The calculated higher order corrections are
relatively small  and  their resummation shows that relativistic expansion
converges sufficiently well. The numerical effect also depends on the model of the nucleon LCDA  however this dependence  does not change the  main qualitative conclusions about the large relativistic corrections. 
     
A description of  excited state $\psi^{\prime}$  is more challenging because
of  relatively large value of   $\left\langle \boldsymbol{v}^{2}\right\rangle
_{\psi^{\prime}}$. In this case the numerical effect from relativistic corrections  is much larger than the
LO contribution  and  is more sensitive to the  higher order contributions.  Therefore a description of
 $\psi^{\prime}$ baryonic decays  suffers from large uncertainties,  which are associated  with the higher-order contributions of the relativistic expansions. 

Taking into account these large relativistic corrections, one cannot expect the $12\%$ rule in Eq.(\ref{12pr}) to hold in the general case. The rather good agreement for the nucleon channel is probably an accidental  consequence of different numerical cancellations.    

The large effect from relativistic corrections raises the question about  a
phenomenological description of baryon decays in  the effective field theory
framework.  Various existing phenomenological estimates for  $J/\psi$ decay
width and angular behaviour  are based on the LO approximation and provide a
qualitative reliable  estimates \cite{Kivel:2022fzk}.  However such  qualitative picture
is violated  by the large value of  relativistic corrections.  The large
negative correction almost cancel the LO contribution, which greatly reduces
the amplitude value  and, therefore, makes the description problematic.  A
possible solution of this problem might be associated with the large and
positive  NLO radiative correction, which is not yet known. Cancellation of
radiative and relativistic  corrections could  resolve  the situation.
 The  large  NLO radiative corrections are already observed in the
exclusive production $e^{+}e^{-}\rightarrow J/\psi\eta_{c}$, see {\it e.g.}  Ref.\cite{Zhang:2005cha}.
 Probably a similar  situation  also arises  in the baryonic decays.
Therefore, the  calculation of the NLO radiative correction   is  necessary in
order to better understand the hadronic decay dynamics. 

\section*{Acknowledgements}
I am grateful to A.~Vairo and  V.~Shtabovenko  for very useful   discussions.  This work was supported by the Deutsche Forschungsgemeinschaft (DFG, German Research Foundation) Project-ID 445769443. 

\section*{\Huge Appendix}
\appendix
\numberwithin{equation}{section}
\setcounter{equation}{0}
\section{Definition and properties of nucleon LCDA }
\label{LCDA}

In this Appendix we provide  definition of the light-cone  matrix element for
 proton  state. The formulae for the antiproton state can be obtained by charge conjugation. 
In order to simplify  formulas, we  also use  the light-cone gauge
\begin{equation}
n\cdot A(x)=0.
\end{equation}

In this case the light-cone three-quark operator which we need for the proton
state can be defined as ($i,j,k$ are the colour indices)
\begin{equation}
O_{\text{tw3}}=\varepsilon^{ijk}\xi_{\alpha}^{i}\left(  z_{1-}\right)
\ \xi_{\beta}^{j}(z_{2-})\ \xi_{\sigma}^{k}(z_{3-}),
\end{equation}
where the projected quark field $\xi(z)\equiv\nbn\psi(z)$, 
the arguments of the fields read
\begin{equation}
z_{i-}=\left(  \bar{n}\cdot z_{i}\right)  \frac{n}{2}.
\end{equation}
The proton flavour structure implies that $O_{\text{tw3}}=u\ u\ d.$

The  twist-3 light-cone  matrix element is  defined as
\begin{align}
&  \left\langle 0\left\vert O_{\text{tw3}}(z_{1},z_{2},z_{3})\right\vert
p(k)\right\rangle =\frac{1}{4}~\left[  \Dsl k~C\right]  _{\alpha\beta}\left[
\gamma_{5}N_{\bar{n}}\right]  _{\sigma}~\text{FT}\left[  V_{1}(y_{i})\right]
\nonumber\\
&  \mskip100mu+\frac{1}{4}\left[  \Dsl k\gamma_{5}C\right]  _{\alpha\beta
}\left[  N_{\bar{n}}\right]  _{\sigma}~\text{FT}\left[  A_{1}(y_{i})\right]
+\frac{1}{4}~k^{\nu}\left[  i\sigma_{\mu\nu}C\right]  _{\alpha\beta}\left[
\gamma_{\bot}^{\mu}\gamma_{5}N_{\bar{n}}\right]  _{\sigma}\text{FT}\left[
T_{1}(y_{i})\right]  ,\label{tw3me}%
\end{align}
where the projected nucleon spinor reads
\begin{equation}
N_{\bar{n}}=\nbn N(k),
\end{equation}
and $C$ is charge conjugation matrix. \ The Fourier transformation
\textquotedblleft FT\textquotedblright\ is defined as
\begin{equation}
\text{FT}\left[  F(y_{i})\right]  =\int Dy_{i}~e^{-iy_{1}k_{-}z_{1+}%
/2-iy_{2}k_{-}z_{2+}/2-iy_{3}k_{-}z_{3+}/2}F(y_{1,}y_{2},y_{3}),
\end{equation}
with the integration measure
\begin{equation}
Dy_{i}=dy_{1}dy_{2}dy_{3}\delta(1-y_{1}-y_{2}-y_{3}).
\end{equation}
Three LCDAs $V_{1}$, $A_{1}$ and $T_{1}$ satisfy the following properties \cite{Braun:2000kw}
\begin{equation}
V_{1}(y_{2},y_{1},y_{3})=V_{1}(y_{1},y_{2},y_{3}),
\end{equation}%
\begin{equation}
A_{1}(y_{2},y_{1},y_{3})=-A_{1}(y_{1},y_{2},y_{3}),
\end{equation}%
\begin{equation}
T_{1}(y_{2},y_{1},y_{3})=T_{1}(y_{1},y_{2},y_{3}).
\end{equation}
The isospin symmetry allows one to get  the following relation, 
see details in Refs. \cite{Braun:2000kw, Chernyak:1984bm}%
\begin{equation}
T_{1}(y_{1},y_{2},y_{3})=\frac{1}{2}\left(  V_{1}-A_{1}\right)  (y_{1}%
,y_{3},y_{2})+\frac{1}{2}\left(  V_{1}-A_{1}\right)  (y_{2},y_{3},y_{1}).
\end{equation}
It is convenient to define the following combination%
\begin{equation}
f_{N}\ \varphi_{3}(y_{1},y_{2},y_{3})=V_{1}(y_{1},y_{2},y_{3})-A_{1}%
(y_{1},y_{2},y_{3}),
\end{equation}
which allows one to describe the set of three LCDAs $V_{1}$, $A_{1}$ and
$T_{1}$ in terms of the one function. The coupling $f_{N}$ describes the normalisation so
that $\varphi_{3}$ is dimensionless and normalised as%
\begin{equation}
\int Dy_{i}\text{ }\varphi_{3}(y_{1},y_{2},y_{3})=1.
\end{equation}

The LCDA   $\varphi_{3}$ also depends on the factorisation scale,  which is
not shown for simplicity
\begin{equation}
\varphi_{3}(y_{1},y_{2},y_{3})\equiv\varphi_{3}(y_{1},y_{2},y_{3};\mu)\text{.}%
\end{equation}
The evolution properties of this LCDA was  studied in Ref.\cite{Braun:1999te}. The moments in Eq.(\ref{phi3ABO}) are multiplicatively renormalisable and  can be calculated as
\begin{equation}
\phi_{ij}(\mu)=\phi_{ij}(\mu_{0})\left(  \frac{\alpha_{s}(\mu)}{\alpha_{s}%
(\mu_{0})}\right)  ^{\gamma_{ij}/\beta_{0}},
\end{equation}
where $\beta_{0}=11-2n_{f}/3$ and   $\gamma_{ij}$ are the corresponding
anomalous dimensions%
\begin{equation}
\gamma_{10}=\frac{20}{9},\ \ \gamma_{11}=\frac{8}{3},\ \gamma_{20}=\frac
{32}{9},\ \gamma_{21}=\frac{40}{9},\ \gamma_{22}=\frac{14}{3}.
\end{equation}

\section{Analytical expressions for the coefficients $A_k,\ B_k$ and $C_k$ defined in Eqs.(\ref{def:Ak})-(\ref{def:Ck})}
\label{appXk}
Here we provide the explicit expressions for the coefficients $\left[X_k\right]_{VV,AV,TT}$ defined in  Eq.(\ref{defXk}). These coefficients are functions of the momentum fractions
\bea
\left[X_k\right]_{VV,AV,TT}\equiv \left[X_k\right]_{VV,AV,TT}(x_{123},y_{123}).
\eea
 Below  we also use following short notation 
\bea
\bar{m}=\frac {m_c}E=\frac{1}{\sqrt{1+\bs v^2}}, \ \   C_{ij}=x_i y_j+ x_j y_i. 
\eea

The coefficients $C_k$ can be obtained using the simple relations
\bea
\left[C_k\right]_{VV,TT}(x_{123},y_{123})&=&\left[A_k\right]_{VV,TT}(x_{213},y_{213}),\\
\left[C_k\right]_{AV}(x_{123},y_{123})&=&-\left[A_k\right]_{AV}(x_{213},y_{213}).
\eea
  The other coefficients read
\bea
\left[A_0\right]_{VV}&=& 2C_{13} - (1-\mb )( x_3 + y_3)-2 \mb(1-\mb),  \\
\left[A_0\right]_{AV}&=&\frac{2}{1+\mb}( - 2C_{13} +(1- \mb^2) ( x_1 - x_2 +  y_1 - y_2)), \\
\left[A_0\right]_{TT}&=&-4(1-\mb)(\mb+1-x_3-y_3). \\
\left[B_0\right]_{VV}&=&-(1-\mb )(2\mb+x_3+y_3),  \\
\left[B_0\right]_{AV}&=& 2(1-\mb )(x_1 - x_2 +  y_1 - y_2), \\
\left[B_0\right]_{TT}&=& 4(2C_{12}+(1-\mb )( x_3 + y_3)+ \mb^2-1). 
\eea
\bea
\left[A_1\right]_{VV}&=&\frac{\mb}{1+\mb}(4( x_1y_3-x_3y_1)+(1-\mb )(x_3 - x_1 +  y_1 - y_3)),  \\
\left[A_1\right]_{AV}&=&-\left[A_1\right]_{VV},  \\
\left[A_1\right]_{TT}&=&0. \\ 
\left[B_1\right]_{VV}&=&0,  \\
\left[B_1\right]_{AV}&=&0,  \\
\left[B_1\right]_{TT}&=&\frac{4\mb}{1+\mb}(4( x_1y_2-x_2y_1)+(1-\mb )(x_2 - x_1 +  y_1 - y_2)).  
\eea
\bea
\left[A_2\right]_{VV}&=&\frac{\mb^2}{(1+\mb)^2}(2C_{13}+(1+\mb)(2+x_3+y_3)-3(1-\mb^2)),  \\
\left[A_2\right]_{AV}&=&-\frac{\mb^2}{1+\mb}(4(x_1+y_1-1)+2(x_3+y_3)+\mb-1),  \\
\left[A_2\right]_{TT}&=&\frac{4\mb^2}{1+\mb}(1-x_3-y_3+\mb). \\
\left[B_2\right]_{VV}&=& \frac{\mb^2}{1+\mb}(x_3+y_3+2\mb),  \\
\left[B_2\right]_{AV}&=&\frac{2\mb^2}{1+\mb}(x_2-x_1+y_2-y_1) ,  \\
\left[B_2\right]_{TT}&=&\frac{4\mb^2}{(1+\mb)^2}(2C_{12}+(1-\mb )(2-x_3-y_3)-2(1-\mb^2)) .   
\eea
\bea
\left[A_3\right]_{VV}&=&\frac{\mb^3}{(1+\mb)^2}(2(x_1-y_1)+y_3-x_3),  \\
\left[A_3\right]_{AV}&=&0,  \\
\left[A_3\right]_{TT}&=&-\frac{4\mb^3}{(1+\mb)^2}(x_3-y_3). \\
\left[B_3\right]_{VV}&=&\frac{\mb^3}{(1+\mb)^2} (x_1-x_2-y_1+y_2),  \\
\left[B_3\right]_{AV}&=&0,  \\
\left[B_3\right]_{TT}&=&\frac{4\mb^3}{(1+\mb)^2} (x_1-x_2-y_1+y_2).   
\eea
\bea
\left[A_4\right]_{VV}&=&\left[B_4\right]_{VV}=\frac{2\mb^4}{(1+\mb)^2},  \\
\left[A_4\right]_{AV}&=&\left[B_4\right]_{AV}=0,  \\
\left[A_4\right]_{TT}&=&\left[B_4\right]_{TT}=\frac{4\mb^4}{(1+\mb)^2}.  
\eea

\end{document}